\documentclass[%
 reprint,
superscriptaddress,
 amsmath,amssymb,
 aps,
floatfix,
]{revtex4-2}
\usepackage[version=4]{mhchem}
\usepackage{float}
\usepackage[nolist,nohyperlinks]{acronym}
\usepackage{graphicx}
\usepackage{dcolumn}
\usepackage{bm}

\usepackage[section]{placeins}
\begin{document}
\raggedbottom
\preprint{APS/123-QED}

\title{Vanadium-based neutron beam monitor}

\author{V. Maulerova}
\email{vendula.maulerova@ess.eu}
\affiliation{European Spallation Source ESS ERIC, SE-221 Lund Sweden}
\affiliation{Division of Nuclear Physics, Lund University, SE-221 00, Lund, Sweden }
\author{K. Kanaki}
\affiliation{European Spallation Source ESS ERIC, SE-221 Lund Sweden}
\author{P. M. Kadletz}
\affiliation{European Spallation Source ESS ERIC, SE-221 Lund Sweden}
\author{R. Woracek}
\affiliation{European Spallation Source ESS ERIC, SE-221 Lund Sweden}
\author{T. Wilpert}
\affiliation{Helmholtz-Zentrum Belin (HZB), Hahn-Meitner-Platz 1, 14109 Berlin, Germany}
\author{K. Fissum}
\affiliation{Division of Nuclear Physics, Lund University, SE-221 00, Lund, Sweden }
\author{A. Laloni}
\affiliation{European Spallation Source ESS ERIC, SE-221 Lund Sweden}
\author{N. Mauritzson}
\affiliation{Division of Nuclear Physics, Lund University, SE-221 00, Lund, Sweden }
\author{F. Issa}
\affiliation{European Spallation Source ESS ERIC, SE-221 Lund Sweden}
\author{R. Hall-Wilton}
\affiliation{European Spallation Source ESS ERIC, SE-221 Lund Sweden}
\affiliation{Universit\`{a} degli Studi di Milano-Bicocca, Piazza della Scienza 3, 20126 Milano, Italy}
\date{\today}

\begin{abstract}
A prototype quasi-parasitic thermal neutron beam monitor based on isotropic neutron scattering from a thin natural vanadium foil and standard $^3$He proportional counters is conceptualized, designed, simulated, calibrated, and commissioned. 
The European Spallation Source designed to deliver the highest integrated neutron flux originating from a pulsed source is currently under construction in Lund, Sweden. 
The effort to investigate a vanadium-based neutron beam monitor was triggered by a list of requirements for Beam Monitors permanently placed in the ESS neutron beams in order to provide reliable monitoring at complex beamlines: low attenuation, linear response over a wide range of neutron fluxes, near to constant efficiency for neutron wavelengths in a range of 0.6-10~\AA{}, calibration stability and the possibility to place the system in vacuum are all desirable characteristics.
The scattering-based prototype, employing a natural vanadium foil and standard \ce{^3He} proportional counters, was investigated at the V17 and V20 neutron beamlines of the Helmholtz-Zentrum in Berlin, Germany, in several different geometrical configurations of the \ce{^3He} proportional counters around the foil. Response linearity is successfully demonstrated for foil thicknesses ranging from 0.04~mm to 3.15~mm. Attenuation lower than 1\% for thermal neutrons is demonstrated for the 0.04~mm and 0.125~mm foils. The geometries used for the experiment were simulated allowing for absolute flux calibration and establishing the possible range of efficiencies for various designs of the prototype. The operational flux limits for the beam monitor prototype were established as a dependency of the background radiation and prototype geometry. The herein demonstrated prototype monitors can be employed for neutron fluxes ranging from $10^3-10^{10}$~n/s/cm$^2$. 
\begin{acronym}
\acro{ESS}{European Spallation Source}
\acro{ERIC}{European Research Infrastructure Consortium}
\acro{DAQ}{Data-acquisition system}
\acro{WFM}{Wavelength Frame Multiplication}
\acro{STF}{Source Testing Facility}
\end{acronym}
\end{abstract}

\maketitle

\section{Introduction}
The \ac{ESS} ERIC \cite{TDRbook,Garoby_2017,ESS_website} is established to design, construct and commission a multi-disciplinary research facility based around a neutron source with unique properties, enabling in turn scientific discoveries in one of the initial 15 instrument stations \cite{ESS_instrument}. The neutron beams of the \ac{ESS}, produced from an accelerated proton beam collisions in a Tungsten target, will be delivered in 2.86 ms long pulses at the frequency of 14~Hz. The neutron flux at its peak will be approximately 30 times higher than the flux from any reactor source  and 5 times higher than the flux from any other currently operational spallation source \cite{TDRbook}. 
\par
More than 150 choppers and numerous guides and other beamline components will adapt the dynamic range and the resolution of the main pulse and guide the neutrons with the desired energies and phase space to each individual instrument station. Due to the high integrated flux, bispectral (thermal and cold neutrons) extraction from the neutron moderators \cite{Bispectral_patent,Butterfly_Pancake_Moderator} and the unique long pulse time structure of the main 2.86 ms pulse from the spallation source, the instruments will be able to use beams with rates and a range of neutron wavelengths that were not available until now and this in turn introduces new possibilities for the experimental research in material sciences \cite{fernandez2017neutron,Schreyer,ESS_science,winkler2006applications}. 
\par
\begin{figure*}
\includegraphics[width=\textwidth]{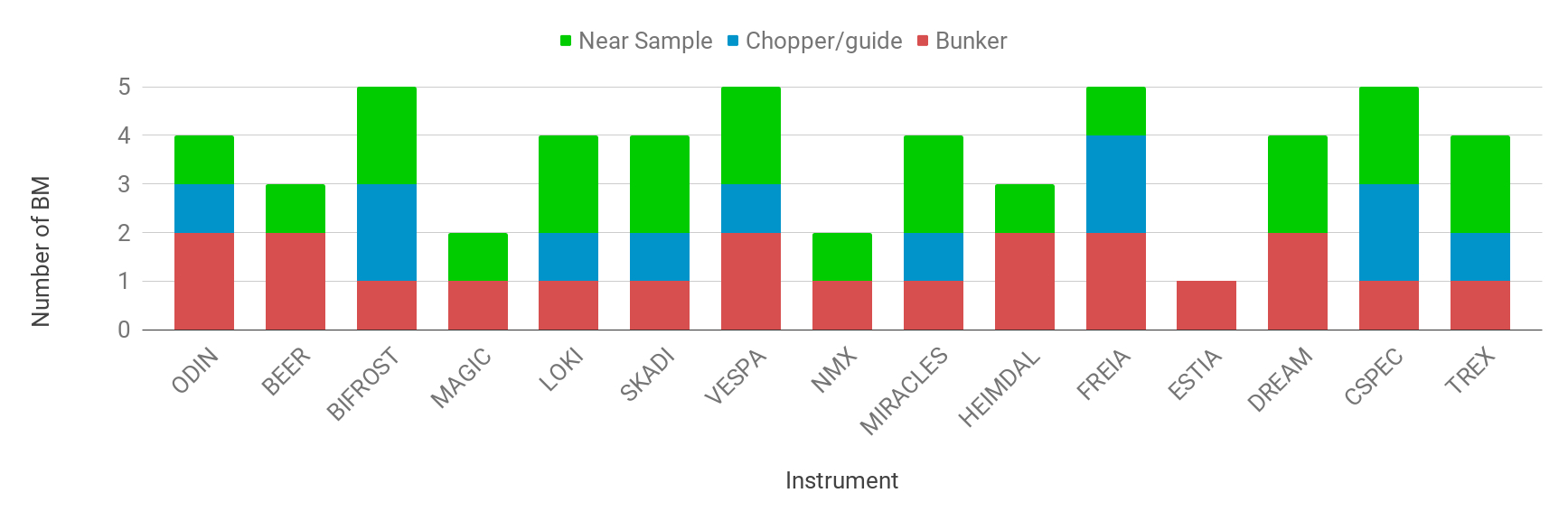}
\caption{\label{fig:BM_perInstrument} The number of Beam Monitors anticipated to be required by the \ac{ESS} instruments as of January 2020. The Beam Monitors are categorised into those near (before or after) the sample, along the neutron guides or next to choppers, or close to the beam extraction from the moderator in the "bunker" region.}
\end{figure*}

The range of wavelengths transported to the instruments sample position runs from 0.6~\AA{} to ca. 20~\AA{}. The lower limit is given by the constraints of neutron transport through curved waveguides, the upper limit by the difficulty in transporting ultracold neutrons. In order to ensure the correct alignment of the beam along the main beam-guides, the proper phase-settings of the choppers and therefore the correct energy range of the neutrons delivered to the instruments, a number of Beam Monitors \cite{FatimaBM} have to be placed along the beam-lines. The multitude of beamline components for instruments at ESS, much higher than for existing neutron sources, implies a greater need for monitoring. Beam Monitors are going to be placed near the monolith, near choppers, near the sample position and in some cases for specific instruments there will be a transmission monitor placed measuring the transmission of the beam through the sample. Figure~\ref{fig:BM_perInstrument} shows the current number of Beam Monitors at various locations for each instrument. Beam monitors that are placed permanently in the neutron beam before the sample position (e.g. after choppers and between guide sections) are required to have low attenuation characteristics. In these locations parasitic and quasi-parasitic methods of beam monitoring that do not disturb the neutron beam are preferred, however such methods do not exist yet to the desired performance.
\par
In this work, a neutron-scattering based prototype Beam Monitor employing a natural vanadium (\ce{^{nat}V}) foil and standard \ce{^3He} detectors in a configuration symmetric around the foil is introduced as a candidate of such a quasi-parasitic Beam Monitor. A vanadium based neutron beam prototype is considered for the locations along the beam-guides and in the near proximity of choppers due to its envisioned low attenuation and high-flux monitoring performance.
\section{Vanadium Beam Monitor principle}
\begin{figure}    
  \includegraphics[width=0.23\textwidth]{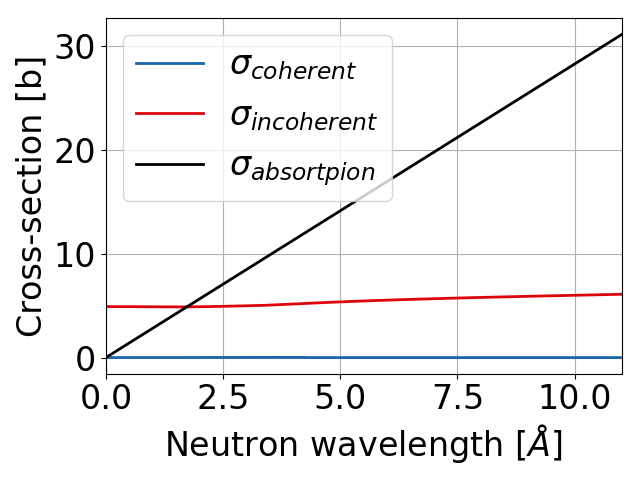}
  \includegraphics[width=0.23\textwidth]{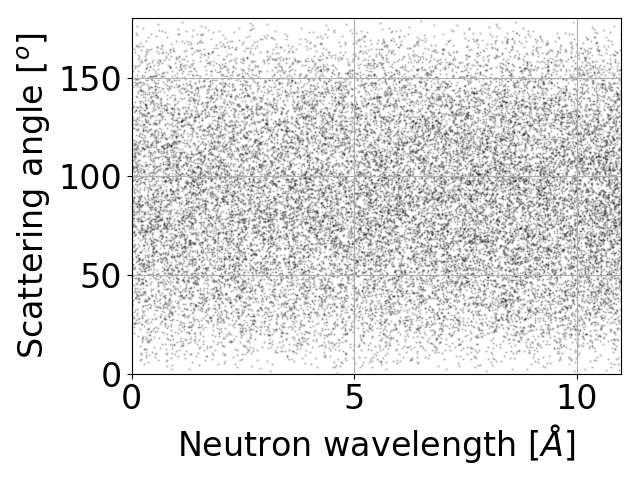}
\caption[Neutron scattering from \ce{^{51}V}.] {Neutron scattering on \ce{^{51}V}, data from NCrystal \cite{NCrystal,NCrystal_wiki}. \ce{^{nat}V} contains 99.75\% of \ce{^{51}V}. Left: Cross section versus wavelength. Right: Scattering angle as a function of wavelength. The lack of features in the scatter plot indicates an isotropic phase space.}
  \label{fig:CrossSectionWavelength}
\end{figure}

To achieve low attenuation of a neutron beam requires to place as little material in the neutron beam as possible. Therefore a neutron-scattering concept is considered where the neutrons scatter from a thin foil and are subsequently detected by commercially available detectors. Further, a foil material with neutron-interaction properties resulting in an easily identifiable signal is highly desirable. Other properties of interest for the base material for a quasi-parasitic Beam Monitor include radiation hardness, that the material isotropically scatters neutrons, good mechanical properties, chemical inertness and optionally vacuum tolerance. Note that few materials scatter thermal neutrons isotropically. In fact besides hydrogen, it is only vanadium that scatters isotropically \cite{IncoherentScattering,VanadiumPhononDist,VibrationSpectra,VanadiumVibrations2,latticeDynamics}. 
Fig.~\ref{fig:CrossSectionWavelength} shows the dependency of the coherent and incoherent scattering as a function of wavelength and the scattering angle as a function of the neutron wavelength range of interest. The coherent scattering is negligible ($\sigma_{coherent}/(\sigma_{coherent}+\sigma_{incoherent})= 0.0036 \approx 0$), while the incoherent scattering is very close to being constant with respect to wavelength, so the number of scattered neutrons is almost independent of wavelength ($< 20$\%). Note that the neutron absorption in vanadium has a linear dependency, so the attenuation changes linearly with the wavelength.
\par
It is possible to also imagine using a non-isotropically scattering material as the base for such a monitor. An example of this might be monitoring the scattering off already existing Al beam window. However, the existence of coherent scattering off this window means that a more detailed and careful calibration and commissioning would be needed. Care would need to be taken in the design to ensure the proportionality of the scattering to flux from the monitor output. Therefore V is chosen here as the material for the monitor.
\par
\par
\begin{figure}   
	\centering
		\includegraphics[width=.5\textwidth]{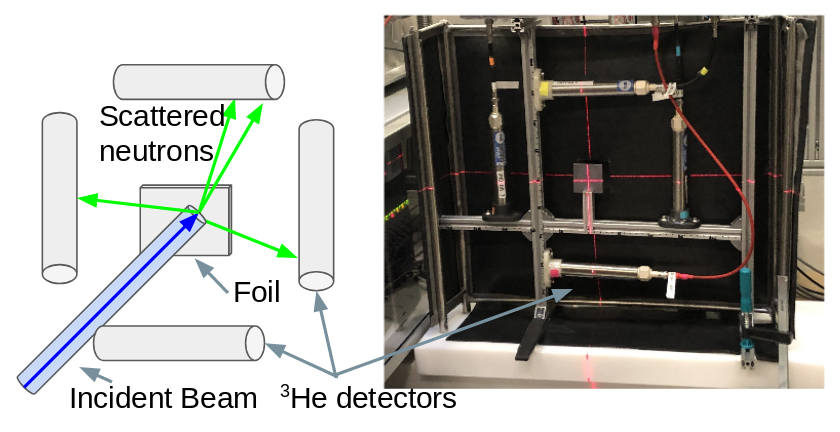}
\caption[Isotropic quasi-parasitic Beam Monitor concept.]{Isotropic quasi-parasitic Beam Monitor concept. The thin foil is placed in the beam of neutrons (denoted by the blue line). Small percentage of the neutrons are scattered from the foil (denoted by the green line). Left: Conceptual cartoon. Right: Photo of the experimental setup for the measurements. The laser cross shows the centre of the neutron beam path.}
\label{fig:BM_Idea}
\end{figure}

The sketch of a basic design of such a Beam Monitor is presented in Fig.~\ref{fig:BM_Idea}. This design has been inspired by a similar device at the ISIS Neutron and Muon Source at the STFC Rutherford Appleton Laboratory near Oxford, UK \cite{ISIS,LET,RobertBewley} on the LET spectrometer. In the figure, the thin foil is placed in a beam and only a low amount of incident neutrons are scattered or captured by the foil. Around the foil, there are symmetrically placed detectors. These are commercially available neutron detectors which record the neutrons scattered from the foil. Additionally, the foil is producing $\gamma$-rays as a product of neutron absorption: these were studied in detail earlier \cite{VM_thesis}. The detectors can be placed symmetrically: in that way a sudden increase of counts in one of the detectors might determine the asymmetry of the beam. For all the measurements carried out with this prototype Beam Monitor, the $^3$He proportional counters from Reuter-Stokes \cite{Reuter-Stokes} were used. The $^3$He proportional counters were 10~cm long tubes, 2.5~cm in diameter. The $^3$He pressure was 8 bar and they were all operated at +1150~V. The distances of the $^3$He proportional counters were varied and 5 different foils with thicknesses depicted in the first column of table \ref{tab:K_coefficients} were used. Boron shielding (3~mm of MirroBor \cite{Mirrotron}) was placed around the \ce{^3He} proportional counters in all directions except that facing directly the foil to reduce background. 
The concept was initially calibrated at the \ac{STF}\cite{STF_FraMe,VM_thesis} in Lund using isotropic Beryllium-based sources \cite{JuliusAmBePuBe}. The induced $\gamma$-ray background was characterized, the measurements were simulated and the viability of the Beam Monitor based on the neutron-scattering was confirmed\cite{VM_thesis}. Thus the proof-of-concept tests were performed at the beamline V17 and the test beamline of the \ac{ESS} V20 at the Helmholtz-Zentum Berlin (HZB), Germany \cite{V17,WORACEK2016102,STROBL201374,Maulerova_2020}. The setup of V17 and V20 is detailed below.

\section{Experimental setup at V17 and V20}
\begin{figure}   
    \centering
    \includegraphics[width=0.5\textwidth]{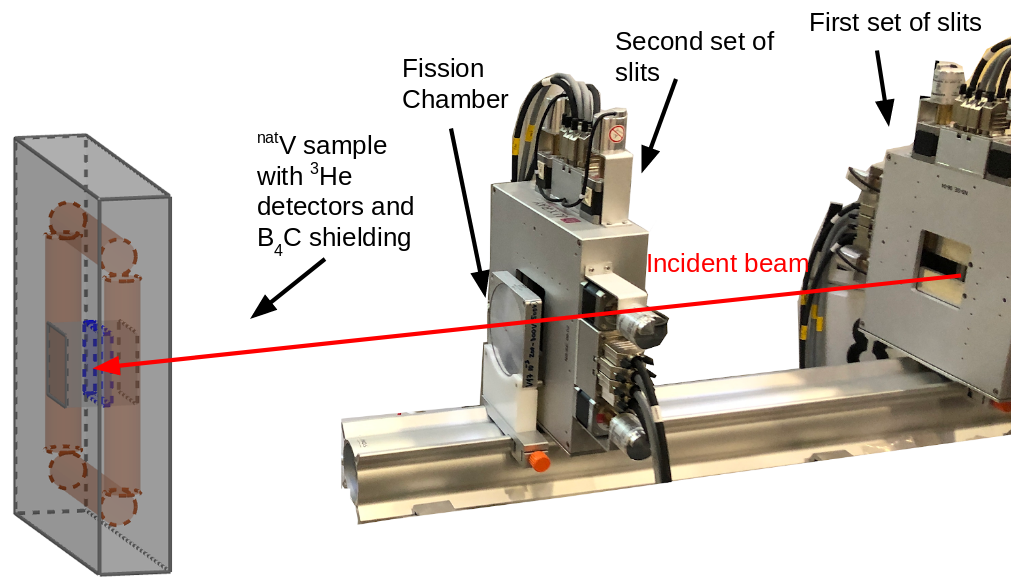}
    \caption[Experimental setup on the beamline at V17.]{Experimental setup on the beamline at V17. The incident beam (red) passes from the right to the left through two sets of (x,y) slits and a fission chamber before striking the \ce{^{nat}V} sample (blue).}
    \label{fig:V17_expSetup}
\end{figure}

Firstly, V17 is an instrument designed to test neutron detectors for different applications. The monochromator (4~mm thick pyrolytic graphite with an area 8~x~1~cm$^2$\cite{V17}) produces a beam of 3.35~\AA{} wavelength. The flux at this beamline was measured to be $2.3 \cdot 10^5$ n/cm$^2$ in 2003 \cite{ThomasWilpert}. Measurements done by other groups in 2019 showed a flux of $1.3 \cdot 10^5$ n/cm$^2$ \cite{LuisMargato}.
Figure~\ref{fig:V17_expSetup} shows the experimental setup for neutron scattering on \ce{^{nat}V} at V17. For the commissioning of the Vanadium Beam Monitor prototype, a reference LND 3053 fission chamber \cite{LND} was placed upstream of the \ce{^{nat}V} foil. The monochromatic neutron beam was collimated using two sets of slits. 
The first set of slits was fixed at 3~$\times$~3~cm$^2$ and the opening between the second set of slits was varied in steps of 2~mm in the x and y direction. This enabled to study the response of the \ce{^3He} detectors which comprised the prototype Beam Monitor as a function of  neutron beam intensity on the \ce{^{nat}V} sample. 
\par
\begin{figure*}   
\centering
  \includegraphics[width=0.8\textwidth]{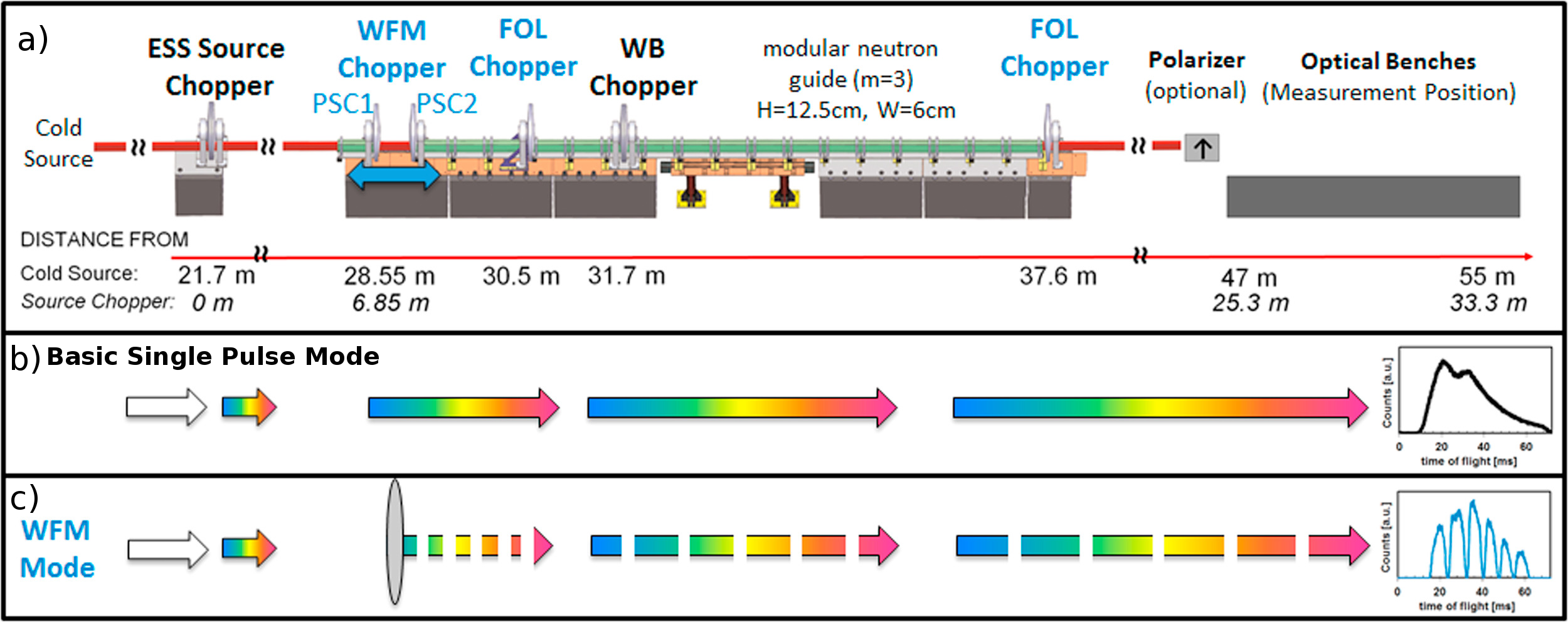}
\caption[Layout of V20.]{Layout of the ESS testbeamline V20. (a) Choppers. The Source Chopper System consists of the ESS Source mimicking Chopper (also capable of producing longer or shorter pulses than the one of the future ESS) and a wavelength-band (WB) chopper. (b) Illustration of the basic single-pulse mode, where the measurement position defines the wavelength (energy) resolution and a continuous spectrum is recorded at the detector. (c) Illustration of the six-fold Wavelength Frame Multiplication (WFM) mode, where the spectrum is divided into sub-frames that are separated in TOF, but overlap in wavelength. Figure  and description from Ref. \cite{WORACEK2016102}. For the herein presented results, the basic single-pulse mode was utilized where the WB chopper was fixed to restrict the wavelength band and the Source Chopper opening was varied in order to adjust the incident neutron flux. The WFM chopper system was parked in open position.}
\label{fig:ChopperSystems}
\end{figure*}

Secondly, the flexible design of V20 (Fig.~\ref{fig:ChopperSystems}) offers convenient control of both the flux and wavelength range of the neutron beam.  
V20 was designed to mimic the ESS pulse structure using two counter-rotating double-disk-chopper systems \cite{STROBL201374}. For the standard operation of V20, the ESS Source Choppers are calibrated to provide 2.86~ms pulse-length with a repetition rate of~14 Hz. This emulates the \ac{ESS}  accelerated-proton beam which is directed onto the ESS tungsten target. Another double-disk-chopper serves as Wavelength-Band chopper in order to prevent frame overlap in time-of-flight, and it can also be used to select more narrow wavelength bands altogether.
After the ESS Source Choppers additional Wavelength-Band (WB) choppers have been used. 
Note that Fig.~\ref{fig:ChopperSystems}c shows also a system of Wavelength Frame Multiplication (WFM) choppers that can be parked in open position if not desired.
\par
\begin{figure}   
    \centering
    \includegraphics[width=0.5\textwidth]{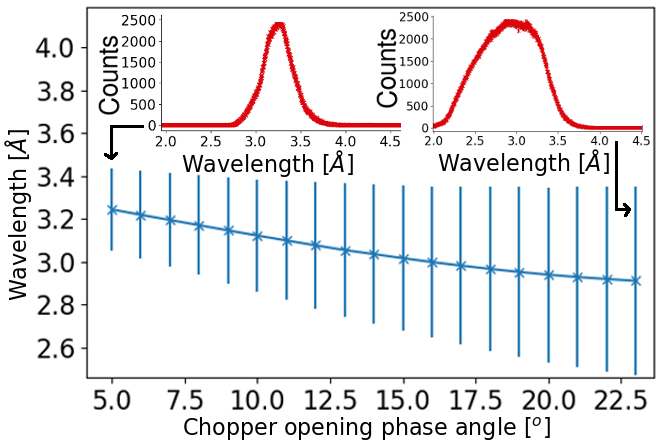}
    \caption[Beamline at V20.]{Wavelength as a function of the chopper 2 opening phase angle. The blue crosses show shows the central wavelength of the wavelength bands recorded by the reference Eurisys Mesures \ce{^3He} Beam Monitor, the vertical blue bars represent the FWHM of the wavelength bands.}
    \label{fig:chopper2}
\end{figure}

In this work, we only utilized the two double-disk-choppers (i.e. the Source Choppers and WB choppers) set to co-rotation and the WFM choppers remained parked in open position. More specifically, the angular opening window of the Source Choppers (from now on referred to as choppers 1 and 2) was varied to produce neutron pulses of different length, hence resulting in different integrated intensities, whereas the angular opening of the WB Choppers (from now on referred as choppers 3 and 4) was fixed throughout the experiment in order to restrict the wavelength band. In this way, we obtained neutron beams with a narrow wavelength around 3.0 \AA{} (this was done in order to compare the monitor efficiency to the one measured at 3.35 \AA{} at V17). By incrementally changing the effective opening of chopper 1 and 2, we were able to study the response of the \ce{^{3}He} detectors in the prototype Beam Monitor as a function of neutron beam intensity on the \ce{^{nat}V} sample, while maintaining the same illumination area. Fig.~\ref{fig:chopper2} shows the change of the width of the wavelength band as a function of the chopper opening phase angle. As the phase of chopper 1 was kept constant the phase of chopper 2 was incrementally increased resulting in Source Chopper opening angles of 5 to 22.5 $^{o}$. With increasing Source Chopper opening angle the FWHM of the wavelength band increases and the centre of the wavelength band shifts as shown in Fig.~\ref{fig:chopper2}. The centre-shift towards smaller wavelengths is a consequence of the WB choppers 3 and 4 kept at constant phase while only delaying the closing of chopper 2, therefore allowing neutrons with higher velocities to traverse the WB choppers.

\begin{figure}   
    \centering
    \includegraphics[width=0.5\textwidth]{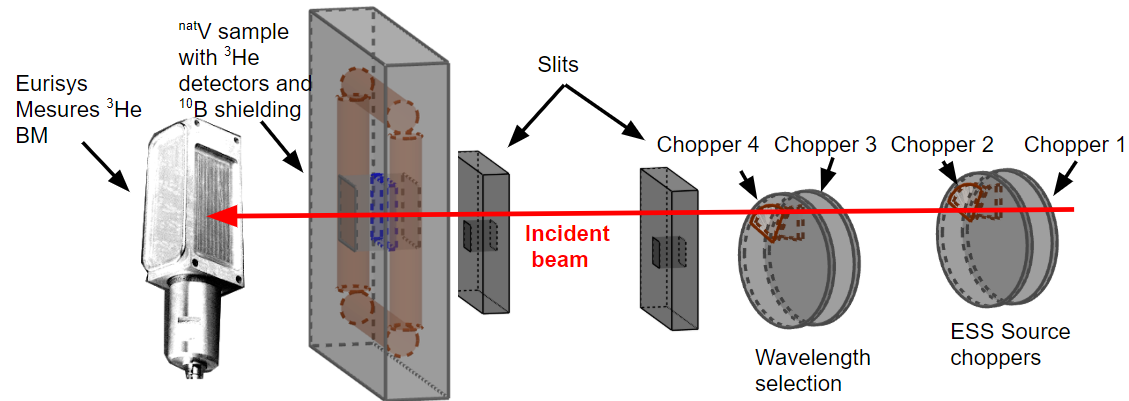}
    \caption[Beamline at V20.]{Experimental setup at V20 beamline. The incident beam (red) passes from right to the left through two sets of choppers and 2 sets of (x,y) slits before striking the \ce{^{nat}V} sample (blue). The slightly attenuated beam is monitored with a a reference Eurisys Mesures \ce{^3He} Beam Monitor.}
    \label{fig:V20_expSetup}
\end{figure}

Fig.~\ref{fig:V20_expSetup} illustrates the experimental setup for neutron scattering on \ce{^{nat}V} at V20. Instead of a fission chamber as used on V17, this time an reference Eurisys Mesures \ce{^3He} beam monitor \cite{He3Monitor} was placed downstream of the \ce{^{nat}V} foil. The neutron beam was collimated using two
sets of slits, each with openings of $3\times 3$~cm$^2$. 
\par
\begin{figure}   
\centering
  \includegraphics[width=0.5\textwidth]{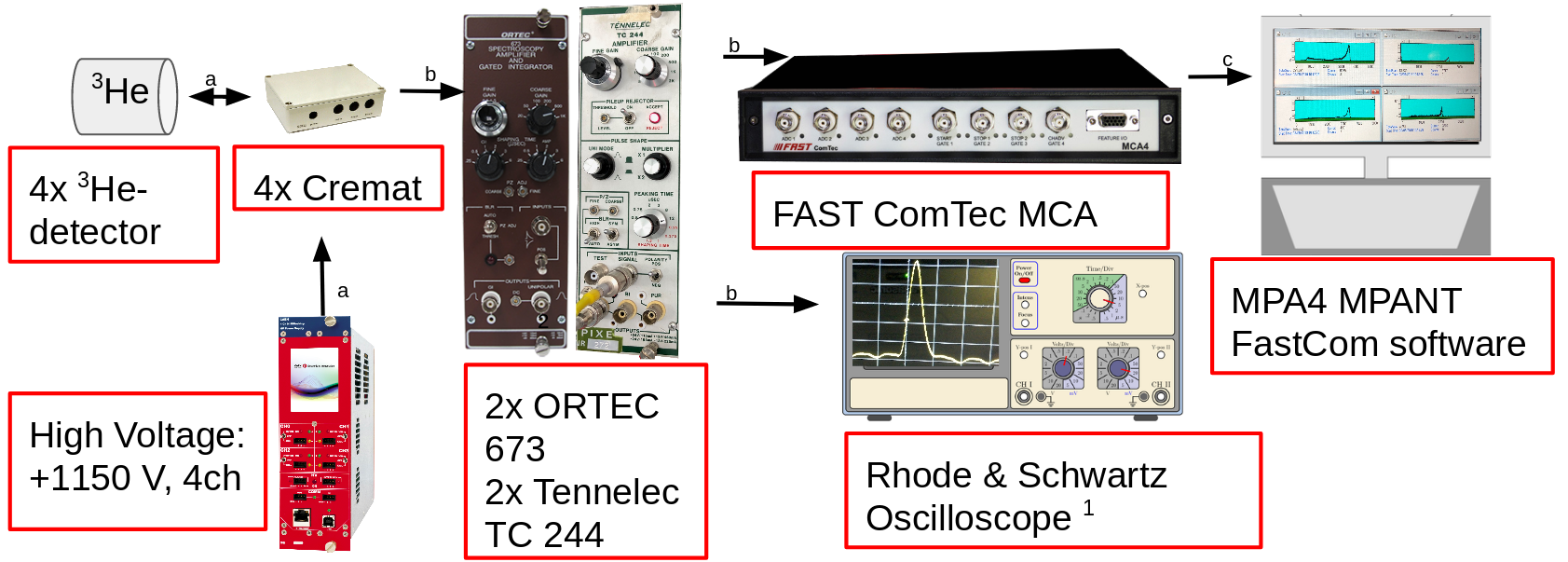}
\caption[DAQ system for the \ce{^3He} detectors.]{DAQ system for the \ce{^3He} detectors. a - HV cable, b - BNC cable, c - MCA-USB link.}
\label{fig:DAQ_Berlin}
\end{figure}

Figure~\ref{fig:DAQ_Berlin} shows the \ac{DAQ} employed for the \ce{^3He} detectors. The electronic chain was common for both beamlines.
Operated at +1150 V, each detector produced a positive current pulse that was passed to a Cremat preamplifier \cite{CREMATPreAmp}. From the preamplifier, two of the signals were passed to NIM ORTEC 673 amplifiers/shapers \cite{ORTEC} while the other two were passed to Tennelec amplifiers/shapers \cite{TC244}. A Rhode \& Schwartz \cite{RohdeSchwartz} oscilloscope was used to visually monitor the pulses periodically. The amplified and shaped pulses were passed to a FAST ComTec MCA4 MCA \cite{FastComMCA}. The MCA integrated the current pulses to yield the charge corresponding to each event and binned these charges resulting in histograms that were stored for subsequent offline analysis.
\section{Flux limits}
\label{sec:Efficiency}
In order to absolutely calibrate the beam: let $I_0$ be the initial intensity of the beam of interest, $I_1$ the number of scattered neutrons from the V-foil in $4\pi$ and $I_2$ is the intensity detected in the detector.
$I_1$ in general depends on the thickness of the foil $d$, the scattering cross section for the neutron scattering in vanadium $\sigma_{s}=5.08$ b (this cross section is mostly independent on the neutron wavelength as seen in Fig.~\ref{fig:CrossSectionWavelength}) and the atomic density $n=7.9 \cdot 10^{22}$~atoms$/$cm$^3$.
The equation is then expressed as:
\begin{equation}
    I_1=n\sigma d I_0=0.36 d I_0.
\end{equation}
Furthermore the relationship between $I_1$ and $I_2$ is a function of solid angle, calculated as the projection of the detector active area on a sphere with the radius $r$ where  $r$ is the distance of the detector from the vanadium foil. The number of counts on the \ce{^3He} counter has to be corrected by the efficiency of the detection:
\begin{equation}
    I_2=\frac{\mathrm{Area}}{4\pi r^2}I_1=\frac{\mathrm{25}}{4\pi r^2}\epsilon_{He3}I_1.
    \end{equation}
From the two equations combined $I_2$ can be expressed as a function of $I_0$:
\begin{equation}
    I_2=\epsilon_{He3}\frac{\mathrm{25}}{4\pi r^2}0.36d\cdot I_0.
\end{equation}
From here then
\begin{equation}
    I_0=\frac{4\pi r^2}{\epsilon_{He3}25\cdot 0.36 d}\cdot I_2=K(r,d) \cdot I_2.
\end{equation}
The coefficient $K(r,d)^{-1}$ is the efficiency of the Vanadium Beam Monitor prototype and $(K(r,d))$ is the absolute calibration constant, required to calculate the original beam intensity $I_0$.
Next to a simple analytical approach, the coefficients $k$ are also calculated by matching the simulations to the measurements. The simulations are performed by using Geant4 \cite{Geant4_1,Geant4_2, Geant4_3}, ESS framework \cite{ESS_framework} and NCrystal \cite{NCrystal,NCrystal_wiki} using the physics list \textit{QGSP\_BIC\_HP}; that way the solid angle is treated more precisely, the crystal structure of vanadium and multiple scattering events are taken into consideration. Also for V17 the changing dimensions of the beam are simulated which would be problematic analytically. 
\begin{table}   
\caption{Analytical and Simulated effect of the \ce{^3He} proportional counter distance on the efficiency and attenuation.}
\label{tab:K_coefficients}
\centering
\begin{tabular}{|c|c|c|c|c|} \hline
d [mm] &  $K_{ana}(r)$  [\%] & $K_{ana}^{-1}$ (r) &  $K_{sim}(r)$  [\%] & $K_{sim}^{-1}$ (r)\\ \hline \hline 
3.15                           & 6.9$r^2$     & 0.144 $\cdot 1/r^2$    & 11$r^2$     & 0.09 $\cdot 1/r^2$         \\
1.00                      & 21.8$r^2$       & 0.045 $\cdot 1/r^2$     & 25.3$r^2$       & 0.039 $\cdot 1/r^2$     \\
0.20                       & 109.0$r^2$        & 0.009 $\cdot 1/r^2$   & 109.1$r^2$        & 0.009 $\cdot 1/r^2$       \\
0.125                  & 174.4$r^2$        & 0.006 $\cdot 1/r^2$   & 196.3$r^2$        & 0.005 $\cdot 1/r^2$         \\
0.04                    & 545.1$r^2$      & 0.002 $\cdot 1/r^2$    & 494.0$r^2$        & 0.002 $\cdot 1/r^2$        \\ \hline
\end{tabular}
\end{table}
Table \ref{tab:K_coefficients} shows the predictions for the efficiency $K(r)^{-1}$ and calibration constants $K(r)$ for different thicknesses as a function of the distance $r$. The simulation model predicts slightly different numbers then the analytical one: this is due to multiple scattering in the vanadium foil when having thicker pieces of vanadium.
\par
In order to estimate limits of the flux with respect to the thickness of the foil $d$, the distance of the center of the counter from the center of the foil and Background rate, the following equations are used.
\begin{align*}
I_{lowest}&>K(r,d)\cdot I_{BG} \\
I_{highest}&<K(r,d)\cdot I_{He3max}
\end{align*}
The highest linearly detectable instantaneous intensity $I_{highest}$ is restrained by the counting limits of the \ce{^3He} proportional counter (from the supplier: $I_{He3max}=$75000~n/s) and the maximum distance $r$ of the \ce{^3He} proportional counters from the center of the foil. 
The lowest detectable instantaneous intensity $I_{lowest}$ is restricted by convolution of the background rate on the \ce{^3He} proportional counter and the distance $r$.
\begin{figure}   
    \centering
    \includegraphics[width=0.5\textwidth]{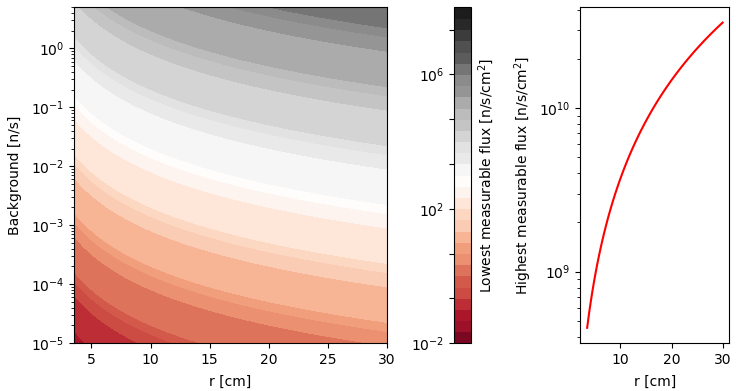}
    \caption{Fluxes restriction as a function of the distance $r$ of the \ce{^3He} proportional counters from the center of the foil and instantaneous background rate (n/s) for 40 $\mu$m thick foil. Left: the lowest detectable flux. Right: the highest linearly detectable flux.}
    \label{fig:Fluxes}
\end{figure}
Fig.~\ref{fig:Fluxes} shows the flux limits in the range of $r \in$ (0,30)~cm. The highest linearly detectable flux is independent of the background, while if the lowest detectable flux changes from vacuum to 1~n/s, in order for the signal to be clearly distinguishable from the background, the required flux can shift by several orders of magnitude.
 In theory there are no restrictions on how far the counters can be placed: GHz range may be achieved for $r=1$m. However this also rises the lowest detectable flux significantly as a function of the background: the higher the background, the higher are the lowest measurable fluxes and at very large values of $r$ there may be very significant systematic influences on the measured flux. So far, it was discussed how the background and the distance of the \ce{^3He} counter affect the count rate capability of the Vanadium Beam Monitor prototype. However there are also other factors: increasing or decreasing the pressure of the counter or changing the detector type can be potentially another tool how to change the flux limits. Another possibility is to make the vanadium foil smaller than the dimensions of the beam. 
\section{Results}
\subsection{V17}
\begin{figure*}   
    \includegraphics[width=1.0\textwidth]{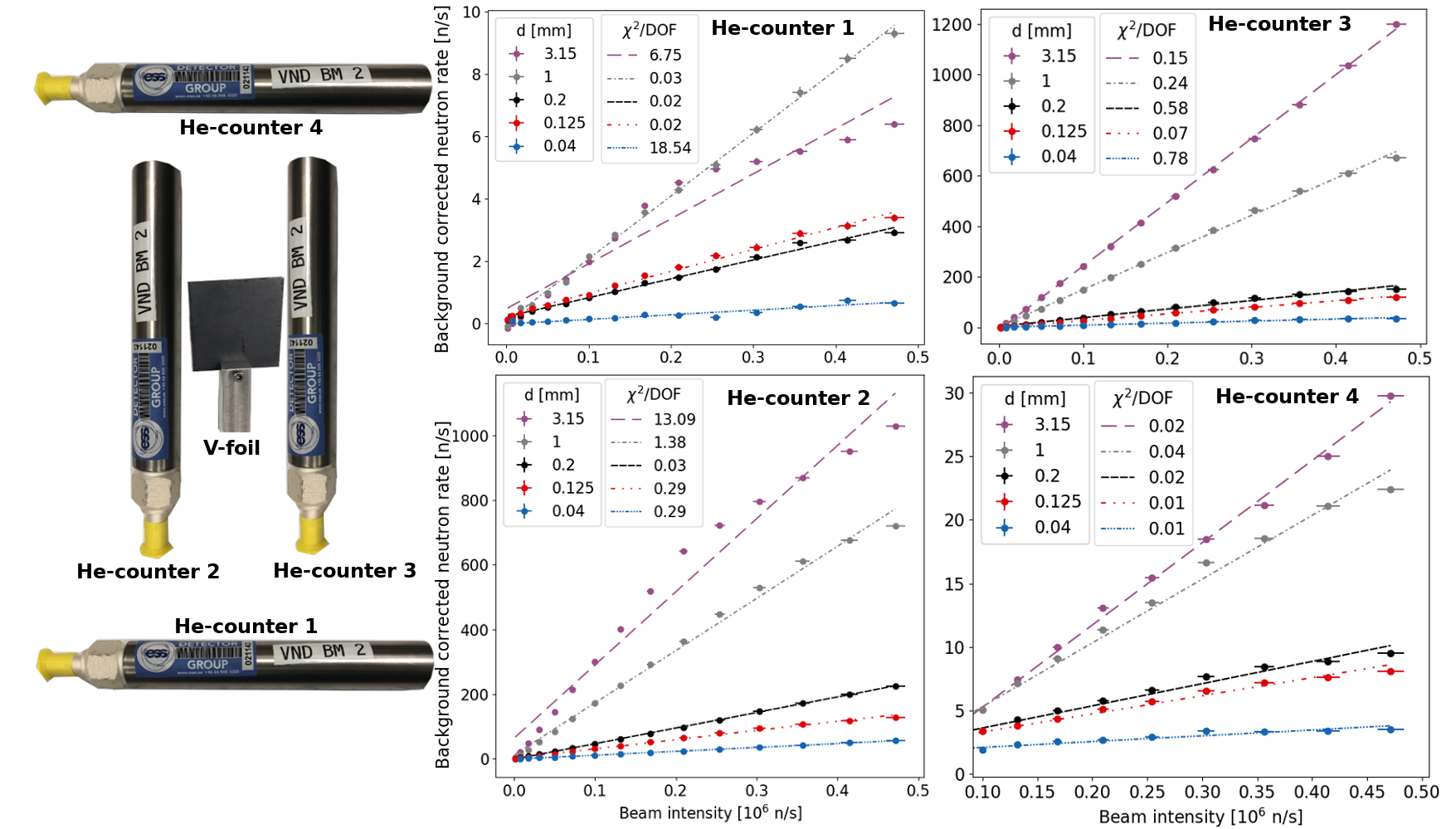}
    \caption[Raw counts from V17.]{\label{fig:V17_rawcounts} Four graphs corresponding to 4 different He-proportional counters placed around the vanadium foil at V17. The efficiency and therefore the flux limitations vary with varying distance of the \ce{^3He} proportional counter from the centre of the vanadium foil.}
\end{figure*}

Fig.~\ref{fig:V17_rawcounts} shows the photo of the geometry of the Vanadium Beam Monitor prototype together with the no-vanadium background corrected count rates recorded by each of the \ce{^3He} proportional counter for the foils with thicknesses of 3.15~mm, 1~mm, 0.2~mm, 0.125~mm and 0.04~mm. Table \ref{tab:DistanceTube} displays the placement of each of the \ce{^3He} proportional counters with respect to the centre of the foil. 
The response is linear with a very good agreement to the fits with $\chi^2/$~DOF(degrees-of-freedom)$<<1$ with several exceptions. The linear inconsistencies for 3.15~mm thick foil when measuring with \ce{^3He} proportional counter 1 and 2 are likely to be explained by the foil being placed asymmetrically, so that the left bottom corner of the foil was not illuminated. This serves as proof of concept for monitoring the asymmetry of the beam. The instability of the 0.04~mm foil for He-counter 1 is due to the low statistics compared to background.
\ce{^3He} proportional counters 2 and 3 were placed near to symmetry to study systematic effects. Shifting the alignment of the counters by 1~mm at 5~cm distance resulted in response different by several hundreds~n/s $\approx$ 0.01 \% of the beam. The further the \ce{^3He} proportional counters are placed, the lower is the rate recorded, the dependence is quadratic. 
The distance $r$ of the $^3$He-counter is a tool of changing the efficiency of the Beam Monitor and therefore changing the highest and lowest flux limits. The efficiency adjustment is a convolution between the thickness of the vanadium foil $d$, the distance of the \ce{^3He} proportional counter $r$.
\begin{table}
\caption{The distances of the \ce{^3He} proportional counters  from the centre of the foil/Beam on V17 and V20}
\label{tab:DistanceTube}
\begin{tabular}{|l|l|l|}
\hline
Counter no. & Beamline & r [cm] \\ \hline \hline
1                             & V20      & 5.2        \\
1                             & V17      & 13         \\
2                             & V17      & 5.1        \\
3                             & V17      & 4.9        \\
4                             & V17      & 11        \\ \hline
\end{tabular}
\end{table}
\par
\begin{figure}   
    \centering
    \includegraphics[width=0.5\textwidth]{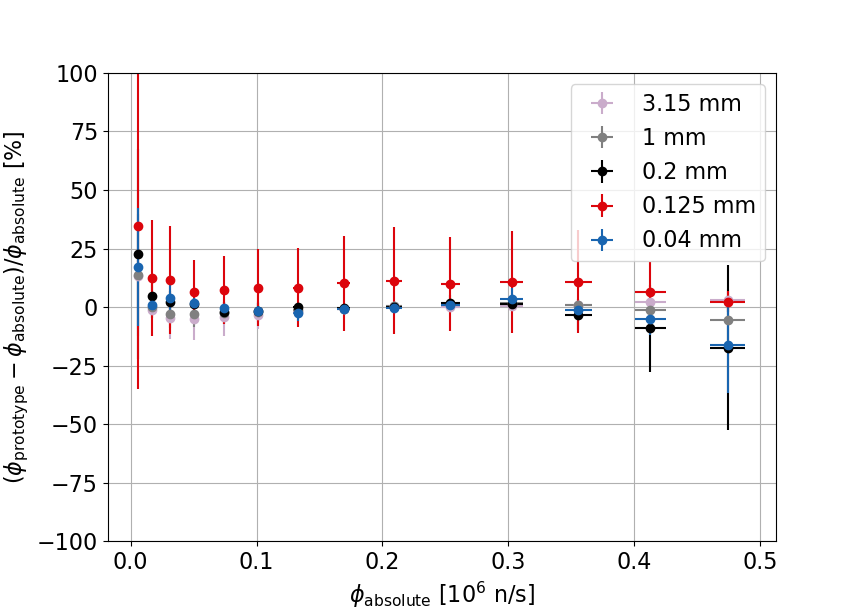}
    \caption[Absolutely calibrated counts from V17.]{Absolutely calibrated counts from V17. The x-axis shows the $\phi_{absolute}$, the absolute beam intensity measured and corrected by the absolutely calibrated 3053 LND fission chamber. The y axis shows the measured values corrected by the results of simulations for each point.}
    \label{fig:V17_absoluteCalibration}
\end{figure}

In order to absolutely calibrate the detector, the experiment had been simulated, see Sec. \ref{sec:Efficiency}.
Fig.~\ref{fig:V17_absoluteCalibration} shows the ratio of the difference between the absolute counts (calculated from the raw measured counts $I_2$ and the factor $K(r,d)$ determined from the simulations) on the prototype Beam Monitor $\phi_{prototype}$ and the efficiency corrected counts on the LND fission chamber $\phi_{absolute}$. The uncertainties in the x direction were estimated by assuming 3\% of variation in the \ce{UO2} coating on the fission chamber. The main component of the uncertainty in the y-direction is asymmetry of the \ce{^3He} proportional counters, and the middle of the beam (and the centre of the foil). The uncertainty was assumed to be 3~mm.  The first value is a low flux related uncertainty, the rest are around 20\%, likely due to the inhomogeneity of the beam. The Beam Monitor shows stable response within the error bars. This demonstrates that for a static system, where the flux is constant, but the illumination area changes, the absolute flux calibration is possible and reliable. It can be seen that for a future implementation with effort understanding the measurement systematics, these uncertainties can be greatly reduced. The downward trend towards the highest $\phi_{absolute}$ values is likely due to the nonlinear beam at high slit opening.
\subsection{V20}
\begin{figure}   
    \centering
    \includegraphics[width=0.5\textwidth]{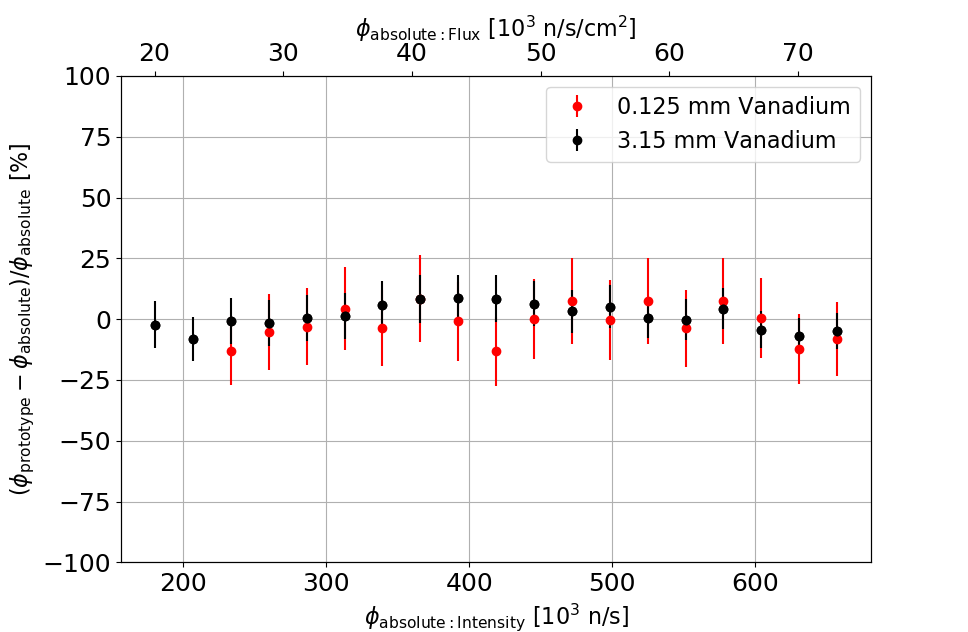}
    \caption[Absolutely calibrated counts from V20.]{Absolutely calibrated counts from the wavelength band at V20. The x-axis shows the $\phi_{absolute}$, the absolute beam intensity measured and corrected by the absolutely calibrated \ce{^3He} Eurisys Mesures detector. The y axis shows the measured values corrected by the results of simulations for each point.}
    \label{fig:V20_absoluteCalibration}
\end{figure}

Fig.~\ref{fig:V20_absoluteCalibration} shows the ratio of the difference between the absolute counts (calculated from the raw measured counts $I_2$ and the factor $K(r,d)$ determined from the simulations) on the prototype Beam Monitor $\phi_{prototype}$ and the efficiency corrected counts on the Eurisys Mesures \ce{^3He} Beam Monitor.
The bottom x-axis shows the neutron beam intensity integrated across the beam, the top x-axis shows the flux.
In order to calibrate the x-axis, the chopper opening angles were linearly plotted as a function of the flux recorded on a reference Eurisys Mesures \ce{^3He} Beam Monitor. The opening angles were then converted to intensity and the flux was calculated by dividing by 9~cm$^2$ of the illumination area.
The absolute calibration is even more stable than in the case of V17, this is expected since a narrower region of fluxes is analyzed and since unlike in the previous case, the illumination area remains the same. The uncertainties are this time around 15\%. The uncertainties are smaller than in case of V17: by maintaining the same illumination area, the setup is less sensitive to the area effects. 
\par
The attenuation in each of the \ce{^{nat}V} foil was measured. This was done by expressing ratio of $I_{\mathrm{foil}}/I_0$ where $I_{\mathrm{foil}}$ is the intensity of the attenuated beam recorded by the Eurisys Mesures \ce{^3He} Beam Monitor for every thickness of the V-foil present in the beam and $I_0$ is the intensity recorded by the Eurisys Mesures \ce{^3He} Beam Monitor when no vanadium was present in the beam. The results are summarized in Table~\ref{tab:V20_attenuation}. 
\begin{table}   
\caption{Measured and expected neutron attenuation in \ce{^{nat}}V at V20.}
\label{tab:V20_attenuation}
\centering
\begin{tabular}{|c|c|c|} \hline
 \ce{^{nat}V} foil thickness $\pm$ 0.01 [mm]  &   \multicolumn{2}{|c|}{Attenuation at 3.1~\AA{} [\%]}  \\ \hline
 ~ & Measured & Expected \\ \hline \hline
3.15                           & 19.09     & 25            \\
1.00                      & 6.91         &9      \\
0.20                       & 1.43    &1.8           \\
0.125                  & 1.11           &1.1    \\
0.04                    & 0.47          &0.4     \\ \hline
\end{tabular}
\end{table}
The suitable foils to use are therefore the 0.125~mm and 0.04~mm foils, because of their exceptionally low attenuation. As demonstrated in Fig.~\ref{fig:Efficiency}, the 0.04~mm foil attenuation remains below 1~\% even for 10~\AA{} neutrons, while typical attenuation of a standard Beam Monitor is in the best case 2\% per mm of window~\cite{FatimaBM}.
\subsection{Wavelength dependency}
\begin{figure}   
    \centering
    \includegraphics[width=0.5\textwidth]{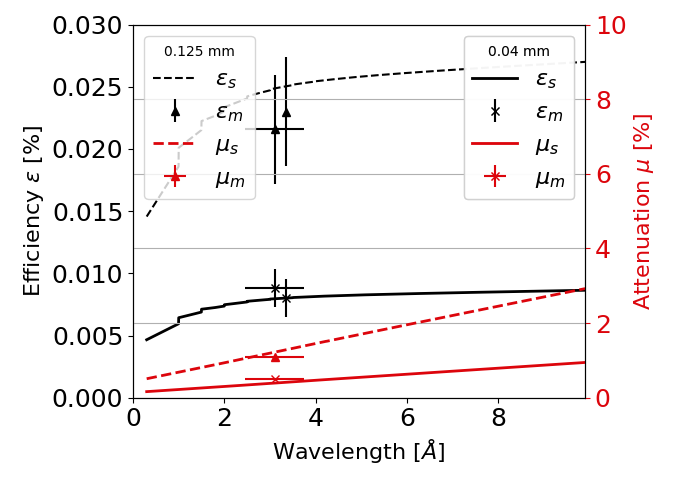}
    \caption[Efficiency and attenuation of the Beam Monitor.]{Efficiency and attenuation of the Beam Monitor prototype. The figure shows the simulated and measured attenuation and efficiency for two of the foils: 0.125~mm (dashed lines, triangle marker) and 0.04~mm (solid lines, cross marker). The left axis (black) displays the measured efficiency ($\epsilon_m$) and the simulated efficiency ($\epsilon_s$) while the right axis (red) displays the measured attenuation ($\mu_m$) and the simulated attenuation ($\mu_s$). The uncertainties in x direction at V20 at 3.1~\AA{} were determined from the FWHM of the reference \ce{^3He} Beam Monitor. For the V17 measurements, the source is a pyrolytic graphite, therefore the uncertainties are not included.}
    \label{fig:Efficiency}
\end{figure}

Fig.~\ref{fig:Efficiency} depicts two of the foils: 0.125~mm and 0.04~mm. The attenuation is simulated for both of them (red axis, red curves, dashed for 0.125~mm and solid for 0.04~mm). The efficiency was also simulated (black axis, black curves, dashed for 0.125~mm and solid for 0.04~mm). The simulation curves are for the case in which the foil is placed 5~cm from the detector: this was the case for experimental setup of one of the \ce{^3He} proportional counters at V17. The experimental point for the V20 efficiency was corrected for the distance (the counter was placed originally placed 6~cm from the centre of the foil). The efficiency is independent of wavelength for higher wavelengths, for lower wavelengths it is determined by the efficiency of the \ce{^3He} proportional counter, which had been taken into consideration in this plot.
Since the coherent scattering does not change with the wavelength, the number of scattered neutrons does not change with the wavelength, making this prototype Beam Monitor suitable to be placed along the beamguides and in the close proximity of the choppers where the wavelengths will be mixed. Attenuation however is a function of the total cross section; the absorption cross section is wavelength dependent. However the instruments with the 10~\AA{} selection can choose the 0.04~mm thin foil, in this case event the attenuation for 10~\AA{} is still below 1 \%.
\par
Using the inverse efficiency ($K$ coefficients from the simulations) enabled to absolutely calibrate the counts recorded at V17.
Fig.~\ref{fig:V17_absoluteCalibration} shows the absolute calibration of the the prototype Beam Monitor at V17 for all the different foil thicknesses. The x-axis are the counts recorded by an 3053 LND fission chamber \cite{LND}. These were corrected by the efficiency given by the supplier (at 1.8~\AA{}) and multiplied by the ratio between the cross section at 3.35~\AA{} and 1.8~\AA{}. This results in an average flux of $5 \cdot 10^4$ n/cm$^2$/s. This is approximately half of the recently measured value by Luis Margato \cite{LuisMargato} $1.3 \cdot 10^5$ n/cm$^2$/s, however this can be explained by differences in methods: Margato's group measured only a little portion of the beam, the inhomogeneity of the beam could cause a flux fluctuation. Furthermore the absolutely corrected flux of the prototype Beam Monitor matches the corrected numbers recorded by the fission chambers, indicating the flux is indeed of this order. 
\subsection{Background induced by the beam}
\begin{figure}   
    \includegraphics[width=0.5\textwidth]{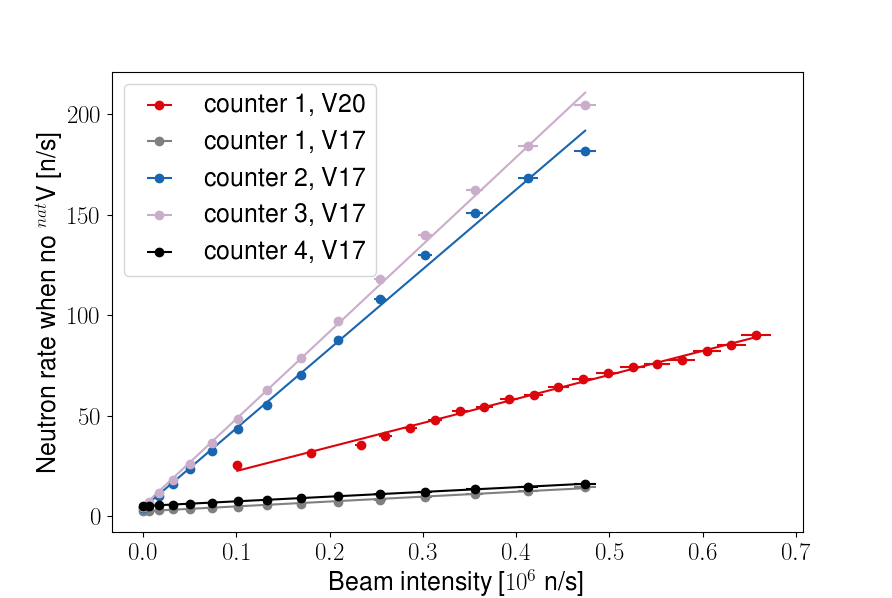}
    \caption{Top: The x-axis shows the absolute beam intensity flowing through through the setup when no vanadium was present. The y axis shows the values recorded by the \ce{^3He} proportional counters. The rate is linear and depends on the distance $r$ of the \ce{^3He} proportional counter from the centre of the beam. Bottom: The x-axis shows the distance $r$ and the y axis shows the slope of the fits from the top picture.}
    \label{fig:Background}
\end{figure}

The focus was to measure the effects of the beam when no vanadium was present in the setup, but the beam was on. In a perfect setup with excellent shielding, where there is no beam dispersion and no other materials in the beam, the response should be the same as the background with the beam off. Fig.~\ref{fig:Background} shows the real response when no vanadium present, but the beam on. The recorded counts in 4 \ce{^3He} counters at V17 (see Table \ref{tab:DistanceTube}) and 1 at V20 have linear resemblance. For all the subsequent analysis, this rate was subtracted to investigate purely the effects of vanadium. It is important for any implementation of such monitors to minimise these effects, as it may not be really linear with respect to the flux in the guide, and so it acts as a systematic uncertainty on flux measurements, and will limit the lower limit of the measurement range of such monitors.
\section{Conclusions}
A prototype quasi-parasitic thermal neutron Beam Monitor employing a \ce{^{nat}V} foil and standard \ce{^3He} proportional counters around the foil had been conceptualized, designed, simulated, calibrated, and commissioned. The cross section for the neutron scattering has been demonstrated as being mainly independent of wavelength, both by simulations and by measurements.  The prototype has been commissioned with beams of neutrons at the V17 and V20 beamlines of the Helmholtz-Zentrum in Berlin, Germany and thus has been proved to function excellently as a quasi-parasitic Beam Monitor. 
\par
This conclusion is based upon the minimal measured beam attenuation with demonstrated linearity over neutron-beam intensities ranging from $10^3$\textendash $5 \cdot 10^6$~n/s. Higher fluxes could be reached by thinner foil, larger distance and lower pressure in the \ce{^3He} proportional counters. 
As variations in the neutron-beam intensity were obtained by both varying the beam-spot size with slits, as well as altering the time structure of the beam with choppers, confidence in the response performance of the Beam Monitor to widely varying beam conditions is high. Absolute calibration has been demonstrated in both V17 and V20 and the background has been characterized.
Based upon desired attenuation being lower than 1\% at 3~\AA{}, the maximum \ce{^{nat}V} foil thickness cannot exceed 0.125~mm, where the attenuation is 1.3\% at 3~\AA{}. These values correspond to the startup commissioning of the ESS instruments. 
\par
As the prototype demonstrated both low neutron-beam attenuation  and response linearity over a reasonable dynamic range, it appears that the concept is sound and worth further detailed design for the ultimate purpose of deployment at ESS.

 \begin{acknowledgments}
We acknowledge the support of Helmholtz-Zentrum Berlin for supporting the beam time at the V20 Testbeamline of the \ac{ESS} and V17 Detector Test Station and the support of Source Testing Facility for supporting the initial characterization tests at Lund University. Furthermore we would like to acknowledge the support from BrightnESS, EU-H2020 grant number: 676548 for personnel and provision of hardware and software. Last, we would like to acknowledge Luis Margato for valuable discussion about the flux at V17. 
 \end{acknowledgments}

\bibliographystyle{unsrt2authabbrvpp}
\bibliography{apssamp}

\end{document}